\documentclass{aa}  

\usepackage{graphicx}
\usepackage{longtable}
\usepackage{hyperref}

\usepackage{txfonts}
\begin{document}

   \title{High-precision analysis of binary stars with planets\thanks{The data presented
   herein were obtained at the W.M. Keck Observatory, which is operated as a scientific partnership
   among the California Institute of Technology, the University of California, and the
   National Aeronautics and Space Administration. The Observatory was made possible by the generous
   financial support of the W.M. Keck Foundation.}
   }

   \subtitle{I. Searching for condensation temperature trends in the HD 106515 system}
   
   \titlerunning{High-precision analysis of binary systems}
   \authorrunning{Saffe et al.}

   \author{C. Saffe\inst{1,2,5}, E. Jofr\'e\inst{3,5,6}, P. Miquelarena\inst{2,5},
    M. Jaque Arancibia\inst{4}, M. Flores\inst{1,2,5},  F. M. L\'opez\inst{1,2,5} \&
  A. Collado\inst{1,2,5}}

\institute{Instituto de Ciencias Astron\'omicas, de la Tierra y del Espacio (ICATE-CONICET), C.C 467, 5400, San Juan, Argentina.
               \email{[csaffe@conicet.gov.ar]}
         \and Universidad Nacional de San Juan (UNSJ), Facultad de Ciencias Exactas, F\'isicas y Naturales (FCEFN), San Juan, Argentina.
         \and Observatorio Astron\'omico de C\'ordoba (OAC), Laprida 854, X5000BGR, C\'ordoba, Argentina.
         \and Departamento de F\'isica y Astronom\'ia, Universidad de La Serena, Av. Cisternas
         1200. 1720236, La Serena, Chile.
        \and Consejo Nacional de Investigaciones Cient\'ificas y T\'ecnicas (CONICET), Argentina
        \and Instituto de Astronom\'ia, Universidad Nacional Aut\'onoma de M\'exico, Ciudad Universitaria, Ciudad de M\'exico, 04510, M\'exico
         }
         
   \date{Received xxx, xxx ; accepted xxxx, xxxx}

 
  \abstract
  {}
   {We explore for the first time the probable chemical signature of planet formation in the remarkable binary
   system HD 106515. The star A hosts a massive long-period planet with {$\sim$9 M$_{Jup}$}
   detected by radial velocity, while there is no planet detected in the B star. 
   We also refine stellar and planetary parameters by using non-solar-scaled opacities when modeling
   the stars.}
   {We carried out a simultaneous determination of stellar parameters and abundances,
   by applying for the first time non-solar-scaled opacities in this binary system,
   in order to reach the highest possible precision.
   We used a line-by-line strictly differential approach, using the Sun and then the A star as reference.
   Stellar parameters were determined by imposing ionization and excitation balance
   of Fe lines, with an updated version of the FUNDPAR program, ATLAS12 model atmospheres and the MOOG code.
   Opacities for an arbitrary composition were calculated through the opacity sampling method.
   The chemical patterns were compared with solar-twins condensation temperature T$_{c}$ trends from literature
   and also mutually between both stars. We take the opportunity to compare and
   discuss the results of the classical solar-scaled method and the high-precision procedure applied here.
   }
   {The stars A and B in the binary system HD 106515 do not seem to be depleted in refractory elements, which is
   different when comparing the Sun with solar-twins.
   Then, the terrestrial planet formation would have been less efficient in the stars of this binary system.
   Together with HD 80606/7, this is the second binary system which does not seem to present a (terrestrial)
   signature of planet formation, and hosting both systems an eccentric giant planet.
   This is in agreement with numerical simulations, where the early dynamical evolution of 
   eccentric giant planets clear out most of the possible terrestrial planets in the inner zone.
   We refined the stellar mass, radius and age for both stars and found a notable difference of $\sim$78 \%
   in $R_{\star}$ compared to previous works. We also refined the planet mass to $m_{p} \sin i =$ 9.08 $\pm$ 0.20 $M_{Jup}$,
   which differs by $\sim$6 \% compared with literature.
   In addition, we showed that the non-solar-scaled solution is not compatible with the classical solar-scaled method,
   and some abundance differences are comparable to NLTE or GCE effects specially when using the Sun as reference.
   Then, we encourage the use of non-solar-scaled opacities in high-precision studies such as the
   detection of T$_{c}$ trends.
   }
   {}
   

   \maketitle
%

\section{Introduction}

In the last years, the achieved high precision in the derivation of stellar parameters
and chemical abundances allowed to study in detail possible differences in stars with and
without planets \citep[e.g. ][]{melendez09,ramirez11,bedell14,saffe17}.
For instance, the search of planet formation or accretion signatures in the photospheric
composition of the stars was performed by looking at the condensation temperature T$_{c}$
trends, with an unprecedented dispersion in metallicity below $\sim$0.01 dex
\citep[e.g. ][]{melendez09,liu14,saffe15,saffe16,saffe17}.
In particular, \citet{melendez09} (hereafter M09) detected a deficiency in refractory elements in the Sun
with respect to 11 solar twins, suggesting that the refractory elements depleted in the
solar photosphere are possibly trapped in terrestrial planets and/or in the cores of
giant planets. The same conclusion was also reached by \citet{ramirez09} and \citet{ramirez10},
using larger samples of solar twins and analogs.

The study of binary or multiple systems plays a central role in the detection of the possible
chemical signature of planet formation, admitting that the stars were born from the same
molecular cloud.
Differential abundances between the components of these systems greatly diminishes
effects such as the Galactic Chemical Evolution (GCE) or the galactic birth place of
the stars, which could affect T$_{c}$ trends \citep[see e.g. ][]{gonzhern13,adi14,adi16}.
In addition, the analysis of two physically similar stars using one of them as
a reference star, allows to diminish the dispersion in both the derivation of stellar
parameters and chemical abundances \citep[e.g. ][]{saffe15}.
Then, a binary system with similar stellar components where only one of them is orbited
by a planet, is an ideal laboratory to look for very small chemical differences 
that could be attributed to the planet formation process.

To date, although more than $\sim$2990 planetary systems are known\footnote{http://exoplanet.eu/catalog/}, 
to find these kind of systems has proven to be a very difficult task.
Examples of these binary systems previously studied in the literature include
16 Cyg, HAT-P-1, HD 80606 and HAT-P-4 \citep[e.g.][]{ramirez11,liu14,saffe15,saffe17}.
There are also binary systems in which circumstellar planets orbit both stars
of the system, such as HD 20781, HD 133131 and WASP-94 \citep{mack14,teske16a,teske16b}.
Then, there is a need for additional stars hosting planets in binary systems to
be compared through a high-precision abundance determination.
Due to their importance, some of these unique systems such as 16 Cyg received the attention of many different
works studying their chemical composition in detail
\citep[e.g.][]{laws-gonzalez01,takeda05,schuler11,ramirez11,tuccimaia14}.
Some works suggested that both stars present the same chemical composition \citep{takeda05,schuler11}
while other studies found that {16 Cyg A} is more metal-rich than the planet host B component
\citep{laws-gonzalez01,ramirez11,tuccimaia14,nissen17}.
In addition, the complete T$_{c}$ trend detected by \citet{tuccimaia14} between the stars of 16 Cyg,
covers a range of only 0.04 dex between the maximum and minimum abundance values of 19 different chemical species
(see their Figure 3). 
More recently, \citet{nissen17} performed a high-precision analysis of this pair using HARPS-N spectra and find
a clear trend with T$_{c}$.
These examples show that the detection of a chemical difference or a possible T$_{c}$ trend as a chemical
signature of planet formation is a challenge, and do require the maximum precision in both stellar parameters
and abundances \citep[for a more complete discussion, see also ][]{saffe18}.
Recently, our group achieved a major step in the pursuit of the highest possible precision. 
For the first time, we used non-solar-scaled opacities in a simultaneous derivation of both stellar parameters
and abundances \citep{saffe18}, for main-sequence and giant stars.
When modelling the atmosphere of the stars, the four stellar parameters
usually taken as (T$_{\rm{eff}}$, {log g}, [Fe/H], v$_{\rm{micro}}$) are now taken
as (T$_{\rm{eff}}$, {log g}, chemical pattern, v$_{\rm{micro}}$).
In this way, we showed that many chemical species show a small but noticeable variation when using the new
doubly-iterated method instead of the usual solar-scaled methods, implying that T$_{c}$ trends could also vary.
Then, we started a new program in order to detect the possible chemical signature of planet formation
in these binary systems, by taking advantage of this improvement in the technique.

\citet{mayor11} first announced through a preprint the detection of a high-mass giant planet
orbiting the star HD 106515 A with a period of $\sim$9.9 yr and a minimum mass
m $sini$ $\sim$10 M$_{Jup}$, as a part of a HARPS radial velocity (RV) survey.
The true mass of this object could correspond to the transition region between planets and brown dwarfs.
The host star belong to a wide binary system together with HD 106515 B, separated by 7.5 arsec
($\sim$250 AU), having both stars a similar mass and a metallicity close to solar \citep{desidera04,desidera06}.
The presence of this massive planet was later confirmed in the work of \citet{desidera12}, who 
performed a RV monitoring of both stars in this system using SARG spectra. However, the authors
do not find significant RV variations on the B star, and rule out additional 
stellar companions by using adaptive optics images.
They propose that the relatively high excentricity of the planet (0.572$\pm$0.011) may arise from the Kozai
mechanism i.e. a dynamical perturbation due to the presence of the wide stellar component.
Then, \citet{marmier13} updated some orbital parameters of the planet by using CORALIE spectra, and proposed 
a possible Kozai mechanism similar to \citet{desidera12}.
We note that this binary system is remarkable for a number of reasons.
First, there is a notable similarity between the stars of this system,
showing an estimated difference in effective temperature and superficial gravity of $+$157$\pm$11 K 
and $-$0.02$\pm$0.15 dex \citep{desidera04}, taken as A $-$ B.
This makes HD 106515 a unique target to analyse through a differential study, belonging to the select
group of binary systems with similar components and having a planet orbiting only one star.
Second, from more than $\sim$2990 planetary systems detected, only 28 of them ($<$ 0.1\%)
are known with a period greater than 9 yr. The long period coverage is very important in order to
properly constrain models of planet formation and migration.
In addition, with a mass higher than 6 M$_{Jup}$, this planet belong to the upper $\sim$15\% of the planetary
mass distribution \citep[see e.g. ][]{marmier13}.
Then, the study of this object give us the possibility, for the first time, to test the possible chemical signature
of planet formation for the case of a high-mass long-period planet.
We take advantage of our recent improvement in the derivation of high-precision abundances \citep{saffe18}
to study this notable binary system with a line-by-line differential approach,
aiming to detect a slight contrast between their components.

This work is organized as follows.
In Section 2 we describe the observations and data reduction, while in Section 3 we present
the stellar parameters, chemical abundance analysis and present a refined value for the planetary mass.
In Section 4 we show the results and
discussion, and finally in Section 5 we highlight our main conclusions.

\section{Observations and data reduction}

Observations of HD 106515 binary system were acquired through the
High Resolution Echelle Spectrometer \citep[HIRES, ][]{vogt94} attached on the right Nasmyth
platform of the Keck 10-meter telescope on Mauna Kea, Hawaii. 
HIRES is a grating cross-dispersed echelle spectrograph, equiped with a
2048x4096 MIT-LL detector with a pixel size of 15 $\mu$m.
The stellar spectra for this work were downloaded from the Keck Observatory
Archive (KOA)\footnote{http://www2.keck.hawaii.edu/koa/koa.html}, under the program ID N158Hr.
The slit used was B2 with a width of 0.574 arcsec, which provides a measured
resolution of $\sim$67000 at $\sim$5200 \AA\footnote{http://www2.keck.hawaii.edu/inst/hires/slitres.html}.

The observations were taken on December, 9th 2013, with the B star observed
immediately after the A star, using the same spectrograph configuration for both objects.
The exposure times were 180 and 240 sec on each target, obtaining a final signal-to-noise
ratio (S/N) of $\sim$300 measured at $\sim$6000 {\AA}.
The final spectral coverage is $\sim$4700-8900 \AA.
The asteroid Iris was also observed with the same spectrograph set-up
achieving a similar S/N, to acquire the solar spectrum useful for
reference in our (initial) differential analysis. We note however that
the final differential study with the highest abundance precision is between
the stars A and B because of their high degree of similarity.

HIRES spectra were reduced using the data reduction package
MAKEE\footnote{http://www.astro.caltech.edu/~tb/makee/}
(MAuna Kea Echelle Extraction), which performs the usual reduction process
including bias subtraction, flat fielding, spectral order extractions, and
wavelength calibration. The continuum normalization and other
operations (such as Doppler correction) were perfomed using
Image Reduction and Analysis Facility (IRAF)\footnote{IRAF is distributed by the National
Optical Astronomical Observatories, which is operated by the Association of Universities for Research
in Astronomy, Inc. under a cooperative agreement with the National Science Foundation.}.

\section{Stellar parameters and chemical abundance analysis}

We derived the fundamental parameters (T$_{\rm{eff}}$, {log g}, chemical pattern, v$_{\rm{micro}}$)
of the stars A and B following the same procedure detailed in our previous work \citep{saffe18}.
We started by measuring the equivalent widths (EW) of Fe I and Fe II lines in the spectra
of our program stars using the IRAF task {\it{splot}}, and then continued with other
chemical species. The lines list and relevant laboratory data were taken from \citet{liu14},
\citet{melendez14}, and then extended with data from \citet{bedell14}, who carefully selected
lines for a high-precision abundance determination.

Stellar parameters and abundances were derived simultaneously, by imposing excitation and ionization
balance of Fe I and Fe II lines.
We used an updated version of the program FUNDPAR \citep{saffe11,saffe18}, which
uses the MOOG code \citep{sneden73} together with ATLAS12 model atmospheres
\citep{kurucz93} to search for the appropriate solution.
The procedure uses explicity calculated (i.e. non-interpolated) plane-parallel local thermodynamic
equilibrium (LTE) Kurucz's model atmospheres, including the internal calculation of
specific opacities through the Opacity Sampling (OS) method.
The first FUNDPAR iterative process searches the iron balance with the usual solar-scaled
model atmospheres. A starting set of parameters and abundances is determined using
EWs and spectral synthesis. Then, the iterative process in FUNDPAR is restarted, but
using ATLAS12 model atmospheres scaled to the last set of abundances found.
This new iteration includes the calculation of specific opacities for the last chemical
pattern specified, and not only a mere change in the abundances of the model.
Thus, ATLAS12 models are described as (T$_{\rm{eff}}$, {log g}, chemical pattern, v$_{\rm{micro}}$)
rather than the usual solar-scaled (T$_{\rm{eff}}$, {log g}, [Fe/H], v$_{\rm{micro}}$).
New stellar parameters and abundances are then successively derived, finishing the
process consistently when the stellar parameters are the same as the previous step
\citep[for more details, see ][]{saffe18}.

Stellar parameters of the stars A and B were determined by applying the
full\footnote{By "full" we mean that line-by-line differences were considered in both the
derivation of stellar parameters and (not only) abundances.} line-by-line differential technique,
using the Sun as standard in an initial approach, and then we recalculate the parameters of the
B star using A as reference.
Firstly, we derived absolute abundances for the Sun using
{5777 K} for T$_{eff}$, {4.44 dex} for {log g} and an initial v$_{turb}$ of {1.0 km s$^{-1}$}. Then,
the solar v$_{turb}$ was estimated by requiring zero slope in the absolute abundances of {Fe I}
lines versus EW$_{r}$ and obtained a final v$_{turb}$ of {0.91 km s$^{-1}$}.
We note however that the exact values are not crucial for our strictly differential study
\citep[see e.g.][]{bedell14,saffe15}.

\begin{figure}
\centering
\includegraphics[width=8cm]{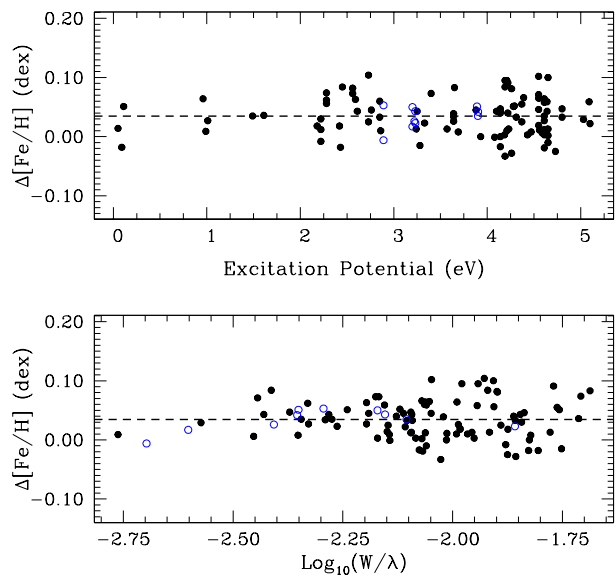}
\caption{Differential abundance vs. excitation potential (upper panel) 
and differential abundance vs. reduced EW (lower panel), for the star A relative to the Sun.
Filled and empty points correspond to Fe I and Fe II, respectively.
The dashed line is a linear fit to the abundance values.}
\label{equil-A-sun}%
\end{figure}

\begin{figure}
\centering
\includegraphics[width=8cm]{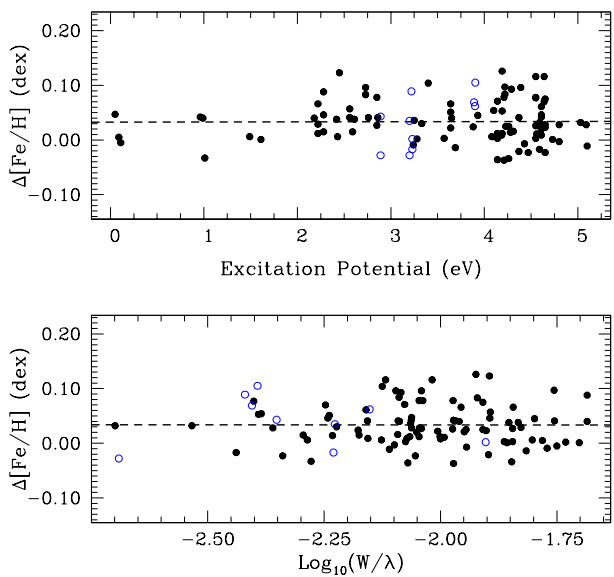}
\caption{Differential abundance vs. excitation potential (upper panel) 
and differential abundance vs. reduced EW (lower panel), for the star B relative to the Sun.
Filled and empty points correspond to Fe I and Fe II, respectively.
The dashed line is a linear fit to the abundance values.}
\label{equil-B-sun}%
\end{figure}

The next step was the determination of stellar parameters of the stars A and B using the
Sun as standard, i.e. (A $-$ Sun) and (B $-$ Sun). The resulting stellar parameters 
for the star A were 
{T$_{eff}$ = 5364$\pm$57 K}, 
{log g = 4.39$\pm$0.18 dex},
{[Fe/H] = +0.016$\pm$0.009 dex,} and
{v$_{turb}$ = 0.79$\pm$0.12 km s$^{-1}$}.
For the star B we obtained
{T$_{eff}$ = 5190$\pm$58 K},
{log g = 4.30$\pm$0.20 dex},
{[Fe/H] = +0.022$\pm$0.010 dex,} and
{v$_{turb}$ = 0.58$\pm$0.15 km s$^{-1}$}.
The errors in the stellar parameters were derived following the procedure detailed in
\citet{saffe15}, which takes into account the individual and the mutual covariance terms
of the error propagation.
We present in Figs. \ref{equil-A-sun} and \ref{equil-B-sun}
abundance vs. excitation potential and abundance vs. reduced equivalent width (EW$_{r}$)
for both stars. Filled and empty points correspond to Fe I and Fe II,  while the dashed lines
are linear fits to the differential abundance values.

\begin{figure}
\centering
\includegraphics[width=8cm]{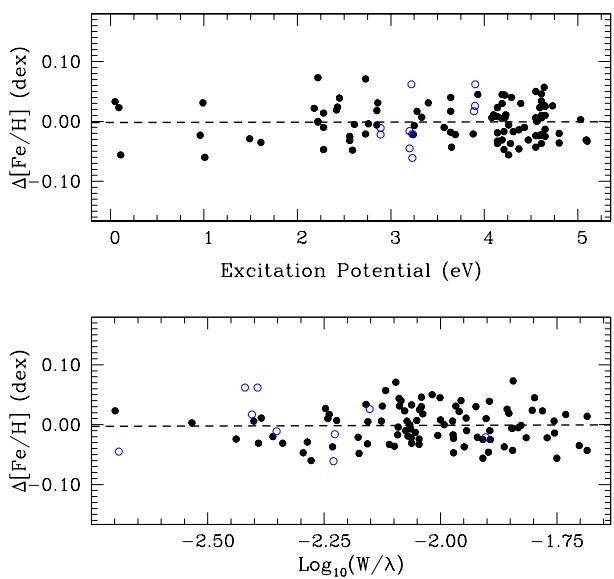}
\caption{Differential abundance vs. excitation potential (upper panel) 
and differential abundance vs. reduced EW (lower panel), for the star B relative to A i.e. (B $-$ A).
Filled and empty points correspond to Fe I and Fe II, respectively.
The dashed line is a linear fit to the data.}
\label{equil-relat}%
\end{figure}

The stellar parameters and abundances of the B star were then redetermined,
but using the A star as reference instead of the Sun, i.e. (B $-$ A) to perform
the differential analysis.
Similar to previous works, we choose the hotter star of the pair as reference
\citep[e.g.][]{saffe15,saffe16,saffe17}.
Figure \ref{equil-relat} shows the plots of abundance vs. excitation
potential and abundance vs. EW$_{r}$, using similar symbols to those used in Figures
\ref{equil-A-sun} and \ref{equil-B-sun}.
The resulting stellar parameters for star B are the same as those obtained 
when we used the Sun as a reference, 
{T$_{eff}$ = 5190$\pm$48 K},
{log g = 4.30$\pm$0.17 dex},
{[Fe/H] = +0.022$\pm$0.009 dex,} and
{v$_{turb}$ = 0.58$\pm$0.12 km s$^{-1}$}.

We computed the individual abundances for the following elements:
\ion{C}{I}, \ion{O}{I}, \ion{Na}{I}, \ion{Mg}{I}, \ion{Al}{I}, \ion{Si}{I}, \ion{S}{I}, \ion{Ca}{I}, 
\ion{Sc}{I}, \ion{Sc}{II}, \ion{Ti}{I}, \ion{Ti}{II}, \ion{V}{I}, \ion{Cr}{I}, \ion{Cr}{II}, \ion{Mn}{I}, 
\ion{Fe}{I}, \ion{Fe}{II}, \ion{Co}{I}, \ion{Ni}{I}, \ion{Cu}{I}, \ion{Sr}{I}, \ion{Y}{II}, and \ion{Ba}{II}.
The \ion{Li}{I} line 6707.8 \AA\ is not present in the spectra.
The hyperfine structure splitting (HFS) was considered for \ion{V}{I}, \ion{Mn}{I}, \ion{Co}{I}, \ion{Cu}{I},
and \ion{Ba}{II} by adopting the HFS constants of \citet{kurucz-bell95} and performing spectral synthesis
with the program MOOG \citep{sneden73} for these species.
The same spectral lines were measured in both stars.
We applied non-local thermodynamic equilibrium (NLTE) corrections to the O I triplet following \citet{ramirez07}. 
The abundances for O I (NLTE) are lower than LTE values ($\sim$0.15 dex and $\sim$0.14 dex for stars A and B).
The forbidden [O I] lines at 6300.31 \AA\ and 6363.77 \AA\ are weak in our stars. 
Both [O I] lines are blended in the solar spectra: with two {N I} lines in the red
wing of [O I] 6300.31 \AA\ and with CN near [O I] 6363.77 \AA\ \citep{lambert78,johansson03,bensby04}.
Then, we prefer to avoid these weak [O I] lines in our calculation and only use the O I triplet.
We also applied NLTE corrections to Ba II following \citet{korotin15}, obtaining NLTE values slightly
lower than LTE ($\sim$0.04 dex and $\sim$0.03 dex for stars A and B), and NLTE corrections to Na
following \citet{shi04}, estimating in this case NLTE values lower than LTE by $\sim$0.05 dex for both stars.
As expected, NLTE corrections resulted very similar for both objects, which is convenient for the differential
analysis.


The final differential abundances [X/Fe]\footnote{We used the standard notation [X/Fe] $=$ [X/H] $-$ [Fe/H]}
for (A $-$ Sun), (B $-$ Sun) and (B $-$ A) are presented in Table \ref{table.abunds}.
Similar to previous works, we present for each specie the observational error $\sigma_{obs}$
(estimated as $\sigma/\sqrt{(n-1)}$ , where $\sigma$ is the standard deviation of the different lines)
as well as systematic errors due to uncertainties in the stellar parameters
$\sigma_{par}$ (by adding quadratically the abundance variation when modifying
the stellar parameters by their uncertainties). For those chemical species with only one line,
we adopted for $\sigma$ the average standard deviation of the other elements.
The total error $\sigma_{TOT}$ was obtained by quadratically adding  $\sigma_{obs}$,  $\sigma_{par}$ and
the error in [Fe/H].

\setcounter{table}{1}
\begin{table*}
\centering
\caption{Differential abundances for the stars A and B relative to the Sun,
and B relative to A i.e. (A $-$ Sun), (B $-$ Sun) and (B $-$ A).
We also present the observational errors $\sigma_{obs}$, errors due to stellar
parameters $\sigma_{par}$, as well as the total error $\sigma_{TOT}$.}
\hskip -0.35in
\scriptsize
\begin{tabular}{ccccccccccccc}
\hline
\hline
  & \multicolumn{4}{c}{(A $-$ Sun)} &  \multicolumn{4}{c}{(B $-$ Sun)} & \multicolumn{4}{c}{(B $-$ A)}\\
Element & [X/Fe] & $\sigma_{obs}$ & $\sigma_{par}$ & $\sigma_{TOT}$ & [X/Fe]  & $\sigma_{obs}$ & $\sigma_{par}$ & $\sigma_{TOT}$ & [X/Fe]  & $\sigma_{obs}$ & $\sigma_{par}$ & $\sigma_{TOT}$  \\
\hline
\ion{C}{I}    &       0.336 &    0.025    &    0.077    &    0.081    &       0.384 &    0.038    &    0.088    &    0.096    &       0.047 &    0.020    &    0.074    &    0.078    \\  
\ion{O}{I}    &      -0.005 &    0.011    &    0.054    &    0.056    &       0.023 &    0.038    &    0.062    &    0.073    &       0.026 &    0.027    &    0.052    &    0.059    \\  
\ion{Na}{I}   &       0.038 &    0.045    &    0.030    &    0.055    &       0.024 &    0.052    &    0.035    &    0.064    &      -0.014 &    0.013    &    0.029    &    0.033    \\  
\ion{Mg}{I}   &       0.105 &    0.016    &    0.025    &    0.031    &       0.086 &    0.024    &    0.025    &    0.036    &      -0.020 &    0.019    &    0.021    &    0.030    \\  
\ion{Al}{I}   &       0.197 &    0.040    &    0.039    &    0.056    &       0.176 &    0.049    &    0.044    &    0.067    &      -0.021 &    0.008    &    0.038    &    0.040    \\  
\ion{Si}{I}   &       0.094 &    0.008    &    0.003    &    0.012    &       0.099 &    0.008    &    0.005    &    0.014    &       0.005 &    0.005    &    0.004    &    0.011    \\  
\ion{S}{I}    &       0.250 &    0.028    &    0.066    &    0.072    &       0.282 &    0.057    &    0.076    &    0.095    &       0.032 &    0.029    &    0.064    &    0.071    \\  
\ion{K}{I}    &      -0.009 &    0.025    &    0.103    &    0.107    &      -0.069 &    0.038    &    0.117    &    0.124    &      -0.061 &    0.020    &    0.098    &    0.101    \\  
\ion{Ca}{I}   &       0.020 &    0.012    &    0.025    &    0.029    &       0.054 &    0.010    &    0.030    &    0.033    &       0.034 &    0.009    &    0.025    &    0.028    \\  
\ion{Sc}{I}   &       0.076 &    0.031    &    0.056    &    0.065    &       0.072 &    0.089    &    0.062    &    0.109    &      -0.004 &    0.058    &    0.051    &    0.078    \\  
\ion{Sc}{II}  &       0.079 &    0.012    &    0.030    &    0.034    &       0.062 &    0.010    &    0.033    &    0.036    &      -0.018 &    0.013    &    0.028    &    0.032    \\  
\ion{Ti}{I}   &       0.143 &    0.012    &    0.014    &    0.021    &       0.164 &    0.013    &    0.016    &    0.023    &       0.021 &    0.008    &    0.014    &    0.018    \\  
\ion{Ti}{II}  &       0.122 &    0.014    &    0.036    &    0.040    &       0.145 &    0.018    &    0.039    &    0.044    &       0.023 &    0.008    &    0.033    &    0.035    \\  
\ion{V}{I}    &       0.162 &    0.016    &    0.025    &    0.031    &       0.155 &    0.022    &    0.029    &    0.038    &      -0.008 &    0.016    &    0.024    &    0.030    \\  
\ion{Cr}{I}   &       0.026 &    0.013    &    0.015    &    0.021    &       0.030 &    0.015    &    0.017    &    0.025    &       0.004 &    0.008    &    0.014    &    0.019    \\  
\ion{Cr}{II}  &      -0.018 &    0.031    &    0.075    &    0.082    &       0.004 &    0.105    &    0.084    &    0.134    &       0.021 &    0.074    &    0.071    &    0.103    \\  
\ion{Mn}{I}   &       0.115 &    0.025    &    0.062    &    0.067    &       0.176 &    0.038    &    0.071    &    0.081    &       0.061 &    0.020    &    0.059    &    0.063    \\  
\ion{Co}{I}   &       0.184 &    0.035    &    0.025    &    0.044    &       0.200 &    0.048    &    0.029    &    0.057    &       0.017 &    0.021    &    0.024    &    0.033    \\  
\ion{Ni}{I}   &       0.031 &    0.006    &    0.004    &    0.011    &       0.040 &    0.007    &    0.004    &    0.013    &       0.008 &    0.005    &    0.003    &    0.011    \\  
\ion{Cu}{I}   &       0.235 &    0.025    &    0.062    &    0.067    &       0.226 &    0.038    &    0.071    &    0.081    &      -0.009 &    0.020    &    0.059    &    0.063    \\  
\ion{Zn}{I}   &       0.208 &    0.041    &    0.027    &    0.050    &       0.209 &    0.025    &    0.031    &    0.041    &       0.000 &    0.016    &    0.026    &    0.032    \\  
\ion{Sr}{I}   &       0.129 &    0.025    &    0.110    &    0.113    &       0.144 &    0.038    &    0.127    &    0.133    &       0.015 &    0.020    &    0.104    &    0.106    \\  
\ion{Y}{II}   &      -0.004 &    0.048    &    0.039    &    0.063    &       0.083 &    0.089    &    0.045    &    0.100    &       0.086 &    0.043    &    0.037    &    0.057    \\  
\ion{Ba}{II}  &      -0.105 &    0.088    &    0.044    &    0.099    &      -0.031 &    0.115    &    0.050    &    0.126    &       0.074 &    0.029    &    0.042    &    0.051    \\  
\ion{La}{II}  &       0.095 &    0.025    &    0.078    &    0.083    &       0.086 &    0.038    &    0.086    &    0.094    &      -0.009 &    0.020    &    0.073    &    0.076    \\  
\ion{Ce}{II}  &      -0.130 &    0.025    &    0.077    &    0.081    &      -0.217 &    0.038    &    0.084    &    0.092    &      -0.088 &    0.020    &    0.073    &    0.076    \\  
\hline
\end{tabular}
\normalsize
\label{table.abunds}
\end{table*}

\subsection{Revised physical stellar and planetary parameters}

From of our new atmospheric parameters in combination with V magnitudes \textit{Gaia} DR2 \citep{gaia18}
parallaxes and stellar evolutionary models, we derived refined stellar mass $M_{\star}$,
radius $R_{\star}$ and age $\tau_{\star}$ for HD106515 A and B. We employed a Bayesian estimation
method and PARSEC isochrones \citep{bressan12} via web interface PARAM 1.3 
\footnote{\url{http://stev.oapd.inaf.it/cgi-bin/param_1.3}} \citep{daSilva06}.
We obtain $M_{\star}$ = 0.888 $\pm$ 0.018 $M_{\odot}$, $R_{\star}$ = 0.910 $\pm$ 0.009 $R_{\odot}$,
$\tau_{\star}$ =  9.233 $\pm$ 2.133 Gyr and $M_{\star}$ = 0.861 $\pm$ 0.015 $M_{\odot}$,
$R_{\star}$ = 0.865 $\pm$ 0.015 $R_{\odot}$, $\tau_{\star}$ =  9.155 $\pm$ 2.199 Gyr for
HD 106515 A and B, respectively. These stellar masses and radii imply stellar densities of
$\rho_{\star}$ =  1.66 $\pm$ 0.05  g cm$^{-3}$ and $\rho_{\star}$ =  1.88 $\pm$ 0.06  g cm$^{-3}$ 
for the A and B component, respectively.
Our estimations of mass are in good agreement with the values derived by \citet{desidera06}, 
who used the same Bayesian method although using isochrones from \citet{girardi00}. Estimations
of radii are no reported in Desidera et al., however, our radii are in perfect agreement with
those provided by \textit{Gaia} DR2.

On the other hand, for HD 106515 A, \citet{marmier13} derived $M_{\star}$ = 0.97 $\pm$ 0.01 $M_{\odot}$
and $R_{\star}$ = 1.62 $\pm$ 0.05 $R_{\odot}$. The masses  agree only within 3$\sigma$, however
their radius is 78\% larger than our estimation. Although they also employed the PARAM code,
but with the stellar models of \citet{girardi00}, we noticed that for this star they reported a
magnitude V = 7.35 taken from the HIPPARCOS catalog \citep{esa97} which is considerably different 
from the one we employed from the Tycho-2 catalog \citep{hog00} and that is displayed on SIMBAD
(V = 7.97). This is probably the main reason for the discrepancies with our stellar parameters, especially radius.

Finally, combining our refined mass estimation of HD 106515 A with the parameters from the spectroscopic orbit 
(velocity semi-amplitude $K$, period $P$, eccentricity $e$) of \citet{marmier13}, we derived an improved 
value of the minimum mass $m_{p} \sin i$ of HD 106515 Ab. Using the equation (1) of \citet{cumming99},
we derive $m_{p} \sin i$ = 9.08 $\pm$ 0.20  $M_{Jup}$, which is $\sim$ 6\% ($\sim$ 175 $M_{\oplus}$)
smaller than the value reported by \citet{marmier13} of $m_{p} \sin i$ = 9.61 $\pm$ 0.14  $M_{Jup}$
\footnote{Value currently reported in The Extrasolar Planets Encyclopaedia.}.
We present in the Table \ref{params} the stellar and planetary parameters derived in this work.

\begin{table}
\centering
\caption{Refined stellar and planetary parameters derived in this work.}
\begin{tabular}{ccc}
\hline
\hline
          & Star A & Star B \\
Stellar Parameters \\
\hline
 T$_{eff}$ [K]                &  5364 $\pm$ 57      & 5190 $\pm$ 58      \\
 log g [dex]                  &  4.39 $\pm$ 0.18    & 4.30 $\pm$ 0.20    \\
 $[$Fe/H$]$ [dex]             & +0.016 $\pm$ 0.009  & +0.022 $\pm$ 0.010 \\
 v$_{turb}$ [km s$^{-1}$]     &  0.79 $\pm$ 0.12    & 0.58 $\pm$ 0.15    \\
 $M_{\star}$ [$M_{\odot}$]    &  0.888 $\pm$ 0.018  & 0.861 $\pm$ 0.015  \\
 $R_{\star}$ [$R_{\odot}$]    &  0.910 $\pm$ 0.009  & 0.865 $\pm$ 0.015  \\
 $\tau_{\star}$ [Gyr]         &  9.233 $\pm$ 2.133  & 9.155 $\pm$ 2.199  \\
 $\rho_{\star}$ [g cm$^{-3}$] &  1.66 $\pm$ 0.05    & 1.88 $\pm$ 0.06    \\
\hline
Planetary Parameters \\
\hline
 $m_{p} \sin i$ [$M_{Jup}$]   & 9.08 $\pm$ 0.20   & \\
\hline
\end{tabular}
\label{params}
\end{table}

\section{Results and discussion}

\begin{figure}
\centering
\includegraphics[width=8cm]{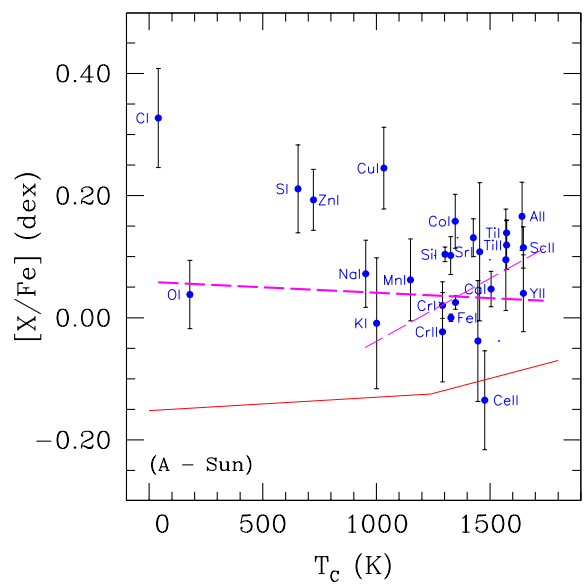}
\caption{Differential abundances {(A $-$ Sun)} vs. condensation temperature T$_{c}$.
Dashed lines are weighted linear fits to all and to refractory species,
while continuous red lines show the solar-twins trend of M09 (vertically shifted for comparison).}
\label{abund-A-sun}%
\end{figure}

\begin{figure}
\centering
\includegraphics[width=8cm]{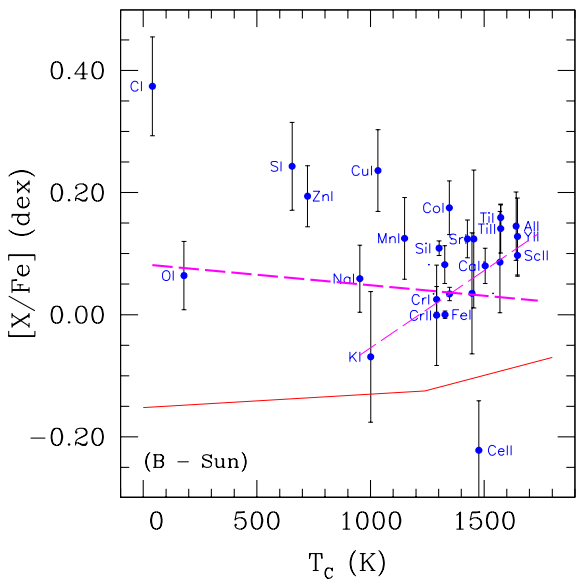}
\caption{Differential abundances {(B $-$ Sun)} vs. condensation temperature T$_{c}$.
Dashed lines are weighted linear fits to all and to refractory species,
while continuous red lines show the solar-twins trend of M09 (vertically shifted for comparison).}
\label{abund-B-sun}%
\end{figure}

The differential abundances of stars A and B relative to the Sun are presented in Figs.
\ref{abund-A-sun} and \ref{abund-B-sun}.
We took the condensation temperatures from the 50\% T$_{c}$ values derived by \citet{lodders03}.
The chemical comparison between one star and the Sun could be affected by GCE effects,
because of their different (chemical) natal environments \citep[see e.g.][and references therein]{tayo14,molla15}.
On the other hand, we discard GCE effects when comparing mutually stars A and B
(owing to their common natal environment), being an important advantage of the differential method.
We corrected GCE effects for (A $-$ Sun) and (B $-$ Sun) by adopting the fitting trends of
\citet{gonzhern13}, with a procedure similar to previous works \citep[e.g. ][]{liu14,saffe15}.
Differential abundances are showed with filled points in Figs. \ref{abund-A-sun} and \ref{abund-B-sun},
while the two dashed lines are weighted linear fits to all abundance values and only to the refractory species.
We used as weight for each chemical element the inverse of the total abundance error $\sigma_{TOT}$.
We also included the solar-twins trend of M09 using a continuous red line,
vertically shifted for comparison.

In this work, we consider as a tentative trend those fits where the slope ranges between 
2-3 $\sigma_{slope}$, and a significant trend when the slope is greater than 3 $\sigma_{slope}$.
The general T$_{c}$ fit to all species showed in Fig. \ref{abund-A-sun} for the A star, present a slightly
negative although non-significant slope (-1.76 $\pm$ 2.67 10$^{-5}$ dex/K) when compared e.g. to the solar-twins trend of M09.
The average abundance of the volatile species (T$_{c}$ $<$ 900 K) is $\sim$0.22 dex,
while the average abundance of the refractory species (T$_{c}$ $>$ 900 K) is $\sim$0.07 dex.
On the other hand, the refractory species taken alone do show a clear positive trend
(slope of +20.6 $\pm$ 4.60 10$^{-5}$ dex/K).
This corresponds to an excess in refractories when compared to the Sun
(which would display a horizontal tendence, not showed) and also when compared to the
solar-twins trend of M09 (red continuous line).
The star B shows in the Fig. \ref{abund-B-sun} a similar behaviour to those showed
by the A star in the Fig. \ref{abund-A-sun} i.e. a slightly negative
although non-significant general slope (-3.46 $\pm$ 3.06 10$^{-5}$ dex/K) together with a positive trend
for refratories (slope of +25.3 $\pm$ 5.29 10$^{-5}$ dex/K).
Then, following a reasoning similar to M09, the stars A and B do not seem to be depleted in
refractory elements when compared to the solar-twins, which is different for the case of the Sun.
In other words, the terrestrial planet formation would have been less efficient in the stars
of this binary system than in the Sun.



\begin{figure}
\centering
\includegraphics[width=8cm]{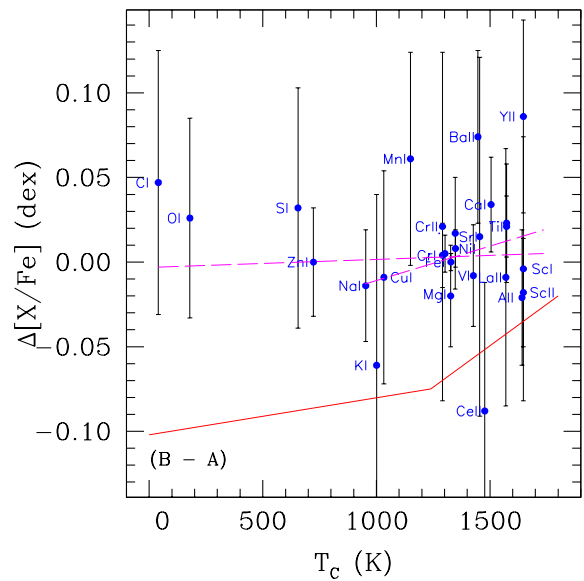}
\caption{Differential abundances (B $-$ A) vs. condensation temperature T$_{c}$.
Long-dashed lines are a weighted linear fit to all and to the refractory species.
The solar-twins trend of M09 is shown with a continuous red line (vertically shifted for comparison).}
\label{relat.tc}%
\end{figure}

The differential abundances of the star B using A as reference i.e. (B - A)
are presented in Fig. \ref{relat.tc}.
This plot corresponds to the abundance values derived with the highest possible precision,
diminishing errors in the calculation of stellar parameters and GCE effects
\citep[e.g. ][]{saffe15}.
Similar to previous Figs., the solar-twins trend of M09 is showed with a continuous red line
(vertically shifted), while the long-dashed lines are weighted linear fits to all and to the refractory
species of (B $-$ A).

The average differential abundances of the volatile species is $\sim$0.026 dex, while the average of the
refractories amount to $\sim$0.005 dex, showing then no clear general trend within the errors
(slope of +0.47 $\pm$ 2.35 10$^{-5}$ dex/K). 
The refractory elements seem to show a positive T$_{c}$ slope (+4.05 $\pm$ 3.86 10$^{-5}$ dex/K),
however there is no significant trend due to the relatively large dispersion of the slope.
We consider that data with higher quality (perhaps higher S/N) is desirable, because there may be
a hidden trend with T$_{c}$ that the current data cannot discern (average error bars of $\sim$0.05 dex).
Following our results, the difference in metallicity for (B $-$ A) is only +0.006 $\pm$ 0.009 dex 
i.e. both stars present almost the same metallicity within our errors.
Then, both stars present very similar metallicity, and there is no clear difference in the relative
content of refractory and volatile elements within our errors.
In other words, there is no clear T$_{c}$ trend between the stars A and B,
and therefore no clear evidence of terrestrial planet formation in this binary system.
Similarly, \citet{liu14} concluded that the presence of a giant planet does not neccesarily
introduce a terrestrial (or rocky) chemical signature in their host stars, by studying the
HAT-P-1 binary system. 
In our case, the massive planet orbiting the A star of the HD 106515 binary system,
present a relatively high eccentricity \citep[0.57, ][]{desidera12,marmier13}.
The presence of long-period planets with eccentric orbits was noted by \citet{marmier13},
who included HD 106515 in their analysis, and attribute the origin of the eccentricity
to a possible Kozai mechanism.
For the case of eccentric giant planets, numerical simulations also found that the
early dynamical evolution of giant planets clear out most of the possible terrestrial planets
in the inner zone \citep{veras-armitage05,veras-armitage06,raymond11}.

\subsection{Solar-scaled vs. non-solar-scaled comparison}

We derived stellar parameters for the stars A and B by applying both
the classical solar-scaled method and the new non-solar-scaled procedure \citep{saffe18}.
The $\Delta$ differences in the parameters taken as (new method $-$ classical method),
amount to $\Delta$T$_{eff}$ +27 K, $\Delta$log g +0.02 dex, $\Delta$[Fe/H] +0.018 dex and $\Delta$v$_{turb}$ +0.07 km s$^{-1}$
for the A star, while for the B star amount to
$\Delta$T$_{eff}$ +23 K, $\Delta$log g +0.02 dex, $\Delta$[Fe/H] +0.012 dex and $\Delta$v$_{turb}$ -0.34 km s$^{-1}$.
Then, for the case of the classical solar-scaled method, we should include in the total error
estimation of these parameters a quantity similar to $\Delta$. In this way,
considering e.g. the T$_{eff}$ of the stars A and B with classical errors of 57 K and 58 K,
and adding quadratically the $\Delta$ differences of 27 K and 23 K, 
would result in a final total error of 63 K and 62 K for stars A and B.
In other words, for the stars of this binary system, the use of the new method
allow a reduction of $\sim$10\% of the total error in T$_{eff}$.

\begin{figure}
\centering
\includegraphics[width=8cm]{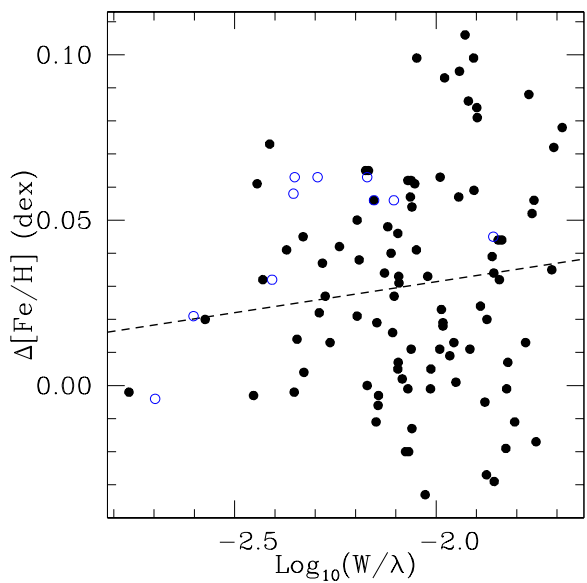}
\caption{Differential abundances (A $-$ Sun) vs. excitation potential using non-solar-scaled opacities,
but forced to the solar-scaled solution.
Filled and empty points correspond to Fe I and Fe II, respectively.
The dashed line is a linear fit to the abundance values.}
\label{out-of-equil}%
\end{figure}

We note that the new method allows an improvement in the calculation of stellar parameters,
while NLTE or GCE corrections only affect the chemical abundances.
We present in the Fig. \ref{out-of-equil} iron abundances vs. EW$_{r}$,
using non-solar-scaled opacities but using the solution found in the solar-scaled method.
Filled and empty points correspond to Fe I and Fe II, respectively, while
the dashed line is a linear fit to the abundance values.
Both the presence of an unbalance in this plot, and the fact that Fe II values are greater
than Fe I values, shows that the solar-scaled solution (requiring e.g. excitation
and ionization balance of iron lines) is not compatible with the new method.
This shows that we can indeed derive a refined solution in stellar parameters
when using non-solar-scaled opacities.

\begin{figure}
\centering
\includegraphics[width=8cm]{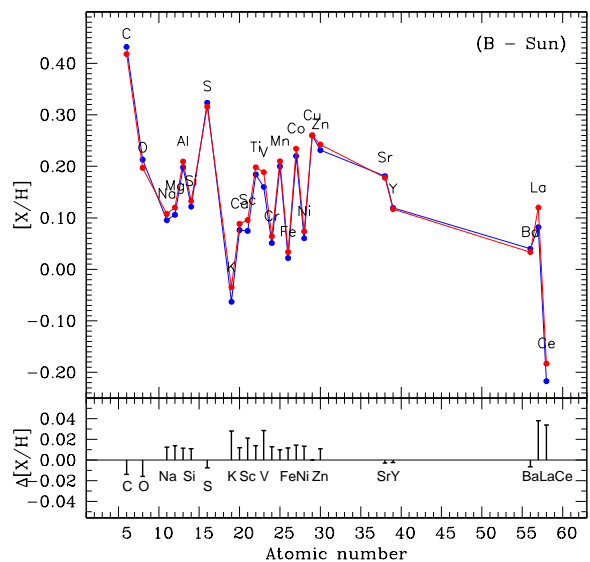}
\caption{Differential abundances (B $-$ Sun) vs. atomic number.
Red circles and lines correspond to the solar-scaled method, while
blue circles and lines corresponds to the new method.
The lower panel presents with bars the difference between both procedures
(as new method $-$ solar-scaled method).}
\label{diff.B}%
\end{figure}

\begin{figure}
\centering
\includegraphics[width=8cm]{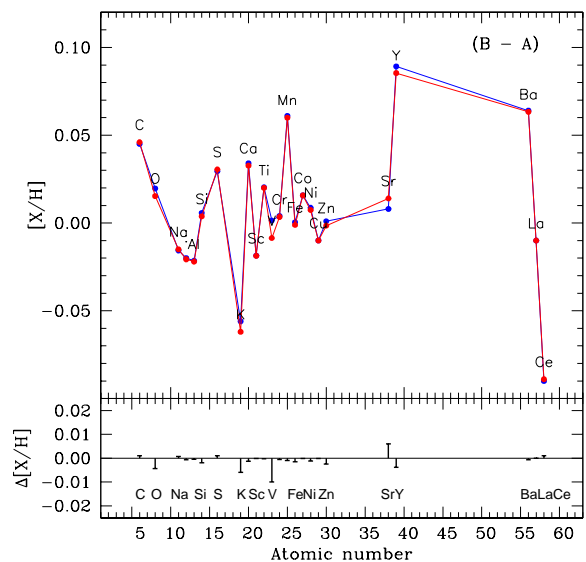}
\caption{Differential abundances (B $-$ A) vs. atomic number.
Red circles and lines correspond to the solar-scaled method, while
blue circles and lines corresponds to the new method.
The lower panel presents with bars the difference between both procedures
(as new method $-$ solar-scaled method).}
\label{diff.BA}%
\end{figure}

We present in the Fig. \ref{diff.B} differential abundances vs atomic number
using the solar-scaled method (blue circles and lines) and
using the new method (red circles and lines) for the star B.
The same plot for the A star shows a similar behaviour.
In the lower panel we present using bars the difference between
both procedures (as new method $-$ solar-scaled method), for each chemical specie.
Some chemical elements are labeled in the plot to facilitate their identification.
Then, the lower panel can be considered as an abundance pattern difference
obtained when using one method or another.
Similar to \citet{saffe18}, we note that C, O and S present lower values
when using the new method, while the rest of the species mostly present 
similar or greater abundance values, by an average of $\sim$0.015 dex.
The greater difference for the star B correspond to La, with a difference
near $\sim$0.04 dex.
We also present in the Fig. \ref{diff.BA} a plot similar to the Fig. \ref{diff.B},
however in this case for (B $-$ A). Due to the physical similarity between stars A and B,
the differences in the individual abundances are usually lower than those of the Fig. \ref{diff.B}.
Most chemical species show almost negligible differences (see lower panel), while
some elements show differences up to $\sim$0.010 dex (for \ion{O}{}, \ion{K}{}, \ion{V}{}, \ion{Sr}{} and \ion{Y}{}).
Then, the differences between both methods are comparable to NLTE corrections or GCE effects
and cannot be easily ignored, specially when comparing the stars using the Sun as reference
rather than the stars between them.

\begin{figure}
\centering
\includegraphics[width=8cm]{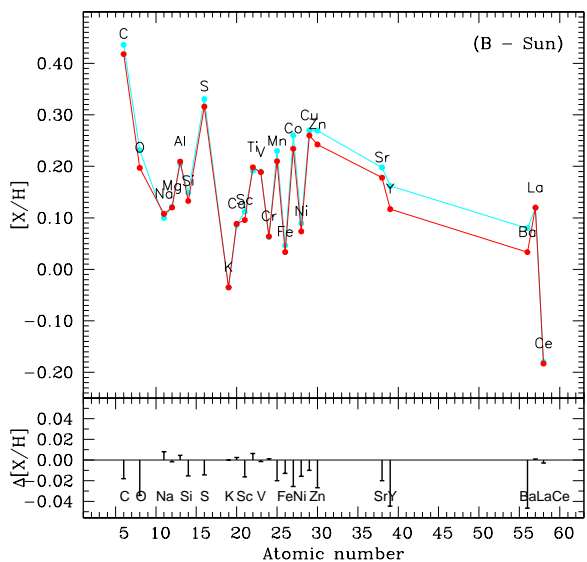}
\caption{Differential abundances (B $-$ Sun) vs. atomic number.
Red circles and lines correspond to the non-solar-scaled method, while
cyan circles and lines correspond to other "corrected" solar-scaled method.
The lower panel presents with bars the difference between both procedures
(as non-solar-scaled $-$  corrected solar-scaled method).}
\label{exp2}%
\end{figure}

We explored the possibility to use "corrected" solar-scaled models rather than the full
non-solar-scaled approach for the star B, by recomputing solar-scaled models but using a modified metallicity.
We used a corrected metallicity similar to the equation 3 of \citet{salaris93}:
$\delta$[Fe/H] $=$ $log_{10}$ (0.638 x $10^{[\alpha/Fe]}$ + 0.362), where $\delta$[Fe/H] is the amount of 
the correction and [$\alpha$/Fe] is the average of the abundances of the alpha elements. 
For the star B we estimated [$\alpha$/Fe] $\sim$ 0.155 dex and $\delta$[Fe/H] $\sim$ 0.104 dex.
Then, the resulting abundances for this correction are presented in the Fig. \ref{exp2},
where red and cyan circles correspond to non-solar-scaled and corrected solar-scaled values.
We note in this plot (see e.g. the lower panel) that there is not a perfect
match between the abundance values derived with the non-solar-scaled and corrected solar-scaled methods.
We suspect that this is due, at least in part, to the fact that solar-scaled models
even with corrected [Fe/H] values, still made use of solar-scaled opacities when deriving
abundances of chemical species other than Fe, giving rise possibly to the small differences
observed in the Fig. \ref{exp2}.


\section{Conclusions}

Aiming to detect a possible chemical signature of planet formation, 
we determined stellar parameters and chemical pattern in both components
of the notable binary system HD 106515, with the highest possible precision.
The star A hosts a massive long-period planet while there is no planet
detected around the star B.
The strong similarity between both stars greatly diminishes errors in the abundance determination,
GCE or evolutionary effects.
We started by deriving the parameters for stars A and B using the Sun as reference i.e. 
(A $-$ Sun) and (B $-$ Sun), and then we recomputed the parameters for the B star using A as
reference i.e. (B $-$ A), obtaining the same results.
We derived very similar chemical patterns for the stars A and B.
By studing the possible temperature condensation T$_{c}$ trends, we concluded that the stars do not
seem to be depleted in refractory elements, which differs from the case of the Sun \citep{melendez09}.
However, we suggest that data with higher quality (perhaps higher S/N) is desirable, because there may be
a hidden trend with T$_{c}$ that the current data cannot discern (average error bars of $\sim$0.05 dex).
Then, following the reasoning of \citet{melendez09}, the terrestrial planet formation would have
been less efficient in the stars of this binary system than in the Sun.
In comparing the stars to each other, the lack of clear T$_{c}$ trend implies that the presence of a
giant planet does not necessarily imprint a (terrestrial) chemical signature on its host star,
similar to previous results \citep{liu14,saffe15}.
We note however that the A star is orbited by a massive eccentric planet, where numerical simulations
found that the early dynamical evolution of giant planets clear out most of the possible
terrestrial planets in the inner zone \citep{veras-armitage05,veras-armitage06,raymond11}.
In this way, both binary systems HD 80606 and HD 106515 do not seem to present a (terrestrial)
signature of planet formation \citep{saffe15}, hosting both systems an eccentric giant planet.

For both stars in the binary system, we refined the stellar mass, radius and age.
In particular, we found a notable difference of $\sim$78 \% in the stellar radius of the HD 106515 A
compared to the value of \citet{marmier13}. This difference would seriously affect the derived planetary
properties (radius, density, etc) of a potential transiting planet that could be detected by the TESS mission
in the next months.
In addition, we refined the minimum planetary
mass to $m_{p} \sin i =$ 9.08 $\pm$ 0.20 $M_{Jup}$, which differs by $\sim$6 \%
when compared with the value obtained by \citet{marmier13}.

We also take the opportunity to compare the parameters derived with non-solar-scaled 
and classical solar-scaled methods. 
We obtained a small but noticeable difference in stellar parameters and individual
chemical patterns. We showed that using non-solar-scaled opacities, the classical solution
cannot verify the standard excitation and ionization balance of iron, similar to \citet{saffe18}.
Also, the difference in abundances between both procedures are comparable to NLTE or GCE effects
specially when using the Sun as reference,
and then cannot be easily avoided in high-precision studies.
Then, we encourage the use of non-solar-scaled opacities in studies which require the
highest possible precision, such as the detection of a possible chemical signature of
planet formation in a binary system.

\begin{acknowledgements}
We thank the anonymous referee for constructive comments that improved the paper.
The authors wish to recognize and acknowledge the very significant cultural role
and reverence that the summit of Mauna Kea has always had within the indigenous
Hawaiian community.  We are most fortunate to have the opportunity to conduct
observations from this mountain. 
M.F. and F.M.L. acknowledge the financial support from CONICET in the form of Post-Doctoral Fellowships.
The authors also thank Drs. R. Kurucz, C. Sneden and L. Girardi for making their codes available to us.

\end{acknowledgements}

\end{document}